\documentclass{aa}
\usepackage{graphicx}
\begin{document}
   \title{Metal enrichment of the intra-cluster medium by thermally and cosmic-ray driven galactic winds}

   \subtitle{An analytical prescription for galactic outflows}

   \author{W. Kapferer \inst{1}
          \and
           T. Kronberger \inst{1}
          \and
           D. Breitschwerdt \inst{2}
          \and
           S. Schindler \inst{1}
          \and
           E. van Kampen \inst{1}
          \and
           S. Kimeswenger \inst{1}
          \and
           W. Domainko \inst{3}
          \and
           M. Mair \inst{1}
          \and
           M. Ruffert \inst{4}
           }

   \offprints{W. Kapferer, \email{wolfgang.e.kapferer@uibk.ac.at}}

   \institute{Institute for Astro- and Particle Physics, University
   of Innsbruck, Technikerstr. 25, 6020 Innsbruck, Austria
   \and
   Zentrum f\"ur Astronomie und Astrophysik, Technische
   Universit\"at Berlin, Hardenbergstrasse 36, 10623 Berlin, Germany
   \and
   Max-Planck-Institut f\"ur Kernphysik, Saupfercheckweg 1, 69117 Heidelberg, Germany
   \and
   School of Mathematics and Maxwell Institute, University of Edinburgh, Edinburgh EH9 3JZ, Scotland, UK}

   \date{-/-}

   \titlerunning{Thermally and cosmic-ray driven galactic winds}


  \abstract
   {}
   {We investigate the efficiency and time-dependence of thermally and cosmic ray driven galactic winds for
   the metal enrichment of the intra-cluster medium (ICM) using a new analytical approximation for the mass outflow. The spatial distribution of the
   metals are studied using radial metallicity profiles and 2D metallicity maps of the model clusters as they would be observed by X-ray telescopes
   like XMM-Newton.}
   {Analytical approximations for the mass loss by galactic winds driven by thermal and cosmic ray pressure are
   derived from the Bernoulli equation and implemented in combined N-body/hydrodynamic cosmological
   simulations with a semi-analytical galaxy formation model. Observable quantities like the mean metallicity, metallicity
   profiles, and 2D metal maps of the model clusters are derived from the simulations.}
   {We find that galactic winds alone cannot account for the observed metallicity of the ICM. At redshift $z=0$ the model clusters have
   metallicities originating from galactic winds which are almost a factor of 10 lower than the observed values. For massive, relaxed clusters we find, as in previous studies, a
   central drop in the metallicity due to a suppression of the galactic winds by the pressure of the ambient ICM. Combining ram-pressure
   stripping and galactic winds we find radial metallicity profiles of the model clusters which agree qualitatively with observed profiles. Only in the
   inner parts of massive clusters the observed profiles are steeper than in the simulations. Also the combination of galactic winds and ram-pressure
   stripping yields too low values for the ICM metallicities. The slope of the redshift evolution of the mean metallicity in the simulations agrees
   reasonably well with recent observations.}
   {}

   \keywords{Galaxies: clusters: general - intergalactic medium - ISM: jets and outflows - Methods: numerical}

   \maketitle
%

\section{Introduction}

A well observable consequence of the interaction processes between
galaxies and the hot, thin intra-cluster medium (ICM) is the high
metal abundance ($\approx$0.3 in solar units) in the intra-cluster
gas. These heavy elements cannot be primordial but must have been
mostly processed in stars within galaxies, and transported to the
ICM by some interaction processes. Galactic winds were one of the
first origins proposed for metal enriched gas found by X-ray
observations (De Young 1978). Recently, it has been shown (Everett
et al. 2008) that the best model to explain the soft X-ray emission
of the Milky Way halo from ROSAT observations is a cosmic ray driven
galactic wind. Further mechanisms that can act as enrichment process
are ram-pressure stripping (Gunn \& Gott 1972), tidal interactions
(Gnedin 1998; Kapferer et al. 2005), active galactic nuclei
(AGNs)(e.g. Moll et al. 2007), and intra-cluster supernovae
(Domainko et al. 2004).

With the first generation of X-ray satellites it was just possible
to measure radial metallicity profiles (see e.g. De Grandi et al.
2004, Boehringer et al. 2005, Baldi et al. 2007, Sato et al. 2009).
Using the current satellites XMM and CHANDRA also the 2D spatial
distribution of the metals became measurable, i.e. metallicity maps
(e.g. Schmidt et al. 2002, Furusho et al. 2003, Sanders et al. 2004,
Fukazawa et al. 2004, Hayakawa et al. 2004). These maps show clearly
a non-uniform and non-spherical distribution of the metals, which
agrees with the results of simulations (e.g. Schindler et al. 2005).
Recent observations also shed some light on the redshift evolution
of the metallicity. Balestra et al. (2007) showed that there is a
mild evolution of the Fe content in the ICM with redshift. From
observations of 56 clusters they conclude that the average Fe
content of the ICM in local clusters is a factor of $\approx$2
larger than at z$\approx$1, even though with a large scatter.
Maughan et al. (2007) confirmed this result with a sample of 115
clusters of galaxies in a redshift interval 0.1 $<$ z $<$ 1.3
observed with Chandra.

Numerical simulations are widely used to investigate the efficiency
and time dependence of the various processes. Several groups have
simulated one or more of the possible enrichment mechanisms of an
isolated galaxy, e.g. ram-pressure stripping (Abadi et al. 1999;
Quilis et al. 2000; Toniazzo \& Schindler 2001; Schulz \& Struck
2001; Vollmer et al. 2001) or galactic winds (e.g. Tenorio-Tagle \&
Munoz-Tunon 1998; Strickland \& Stevens 2000).

On galaxy cluster scales metal enrichment processes are implemented
into cosmological simulations to explain the overall enrichment of
the ICM. De Lucia et al. (2004) and Nagashima et al. (2005) use
combined semi-analytic techniques and N-body simulations to
calculate the overall ICM metallicity. They find that mainly massive
galaxies contribute to the enrichment and that there is a mild metal
evolution since z=1 and a stronger evolution for z$>1$. Tornatore et
al. (2004, 2007) and Valdarnini (2003) use smoothed particle
hydrodynamics (SPH) with detailed yields from type Ia and II
supernovae. Cora (2006) also uses combined N-body/SPH simulations
together with a semi-analytic model. Aguirre et al. (2001)
investigated the efficiency of galactic winds at z$\simeq 3$. They
did, however, not calculate the distribution of the metals that
arise from the winds. They conclude that there are degeneracies in
the results and that other enrichment processes might lead to the
same result.

In Kapferer et al. (2007a) we have simulated, for the first time,
two enrichment processes (galactic winds and ram-pressure stripping)
acting simultaneously. Compared to our previous work (Schindler et
al. 2005; Kapferer et al. 2006; Domainko et al. 2006; Moll et al.
2007) this investigation was extended to high redshifts. With this
approach a reasonable agreement with observed metallicity profiles
was achieved. The wind model adopted in Kapferer et al. (2007a) was,
however, a phenomenological one, in which the mass loss scales
linearly with the star formation rate of the galaxy. This model does
not allow for a deeper insight into the physical processes involved
in the outflow and neglects environmental effects such as the
possible suppression of galactic winds by high ICM pressure
(Schindler et al. 2005). We extend here the work presented in
Kapferer et al. (2007a) by introducing a new analytical model for
the mass loss due to galactic winds. A major advantage of this
implementation of outflows by galactic winds presented in this work
compared to the outflow model in Schindler et al. (2005) is the
analytical nature of the calculation. Whereas in Schindler et al.
(2005) the mass outflow was calculated with an external numerical
scheme (Breitschwerdt et al. 1991) individually for each galaxy at
each timestep, the scheme presented in this work can be easily
implemented on top of semi-analytical models due to its analytical
nature.

The paper is organized as follows: in Sect. 2 we present the new
analytical model used for this work. In Sect. 3 we present the
numerical setup and the implementation of the analytical model. The
results are presented and compared to observations in Sect. 4. We
end with a summary of the main conclusions in Sect. 5.

\section{Analytical wind model}

An analytical model for galactic winds driven by thermal, cosmic
ray, and magnetohydrodynamic (MHD) wave pressure has been presented
by Breitschwerdt et al. (1991). To account for the plane-parallel
geometry of the outflow near the disc and the spherical geometry at
large distances from the disc, Breitschwerdt et al. (1991) use a
flux tube geometry, where the area cross section $A(z)$ varies as

\begin{equation}
A(z)=A_0 \left[1+\left(\frac{z}{Z_0}\right)^2\right].
\end{equation}

\noindent Here $z$ denotes the height above the plane of the galaxy,
$A_0$ the area at a reference level typically 1kpc above the plane
of the galaxy, and $Z_0$ is the height above the galactic plane, at
which flux tube divergence becomes significant, taken to be equal to
the galactic radius here. In this geometry the conservation of mass
and energy of the outflow can be written as

\begin{equation}\label{masscons}
\rho u A=const.
\end{equation}
and
\begin{eqnarray}\label{bernoulli}
\rho u A
\left[\frac{1}{2}u^2+\frac{\gamma_g}{\gamma_g-1}\frac{p_g}{\rho}+\Phi\right]&+&
\frac{\gamma_c}{\gamma_c-1}p_cA(u+v_A)\\
\nonumber &+& p_wA(3u+2v_A)=const.,
\end{eqnarray}

\noindent where $\rho$ is the gas mass density and $u$ the bulk
velocity of the gas. $\gamma_g$ and $\gamma_c$ denote the adiabatic
indices of the thermal plasma and the cosmic rays, respectively. We
assume a non-relativistic monatomic gas, i.e. $\gamma_g=5/3$, while
for the cosmic rays we chose the ultra-relativistic case, i.e.
$\gamma_c=4/3$. $p_g$, $p_c$, and $p_w$ refer to the pressure of the
gas, the cosmic rays, and the MHD waves, respectively. The cosmic
ray pressure is obtained by taking the appropriate moment of the
particle distribution function. Pressure gradients can be calculated
in a hydrodynamic approximation using dynamical equations presented
in Breitschwerdt et al. (1991). Finally, $v_A$ refers to the
Alfv\'{e}n velocity.

Using Eq. \ref{bernoulli}, the Bernoulli equation, we derive an
analytical approximation for the mass loss due to thermally driven
winds. With the continuity equation (Eq. \ref{masscons}) and the
assumptions $p_c \ll p_g$ and $p_w \ll p_g$, Eq. \ref{bernoulli}
simplifies to

\begin{equation}\label{bernoulli_1}
\left[\frac{1}{2}u_0^2+\frac{\gamma_g}{\gamma_g-1}\frac{p_{g0}}{\rho_0}+\Phi_0\right]=
\left[\frac{1}{2}u_{\infty}^2+\frac{\gamma_g}{\gamma_g-1}\frac{p_{g\infty}}{\rho_{\infty}}+\Phi_{\infty}\right],
\end{equation}

\noindent where the subscripts $0$ refer to the quantities at the
reference level, taken to be at $z=1$ kpc. While the subscript
$\infty$ corresponds to the values of the unbound outflow at large
distances, i.e. $z\rightarrow \infty$. For large $z$ the
gravitational potential and the gas pressure tends towards zero.
With these approximations we can further simplify Eq.
\ref{bernoulli_1} to

\begin{equation}\label{bernoulli_2}
\frac{1}{2}u_0^2+\frac{5}{2}\frac{p_{g0}}{\rho_0}+\Phi_0=
\frac{1}{2}u_{\infty}^2.
\end{equation}

\noindent Here we have also used $\gamma_g=5/3$. Eq.
\ref{bernoulli_2} holds if $c_{\infty}\ll u_{\infty}$, which is
always fulfilled for high Mach number winds. It is now possible to
solve for the initial outflow velocity $u_0$:

\begin{equation}\label{bernoulli_3}
u_0=\sqrt{u_{\infty}^2+v_{esc}^2-\frac{5}{\gamma}c_0^2},
\end{equation}

\noindent where we inserted the relation

\begin{equation}\label{crel}
c^2=\gamma\frac{p}{\rho}
\end{equation}

\noindent for the sound speed $c$ and the definition of the escape
velocity $v_{esc}=\sqrt{-2\Phi_0}$. The local mass loss rate per
unit area associated with an outflow with an initial velocity $u_0$
given by Eq. \ref{bernoulli_3} is then

\begin{equation}
\dot{m}=\rho_0u_0.
\end{equation}\label{massloss_eq}

\noindent The total mass loss of a galaxy due to galactic winds can
be found by integrating Eq. \ref{massloss_eq} over the galactic disc
or the area over which mass loss occurs, i.e.

\begin{equation}
\dot{M}=2\times 2\pi \times\int_0^{R_{gal}}\rho_0(r)u_0(r)dr,
\end{equation}\label{massloss_total}

\noindent where R$_{gal}$ is the star formation radius of the
galaxy.

In the same way Breitschwerdt et al. (1991) derived an analytical
approximation for the mass loss due to cosmic ray driven winds, i.e.
the case where $p_g \ll p_c$ and $p_w \ll p_c$, which reads

\begin{equation}
\dot{m}=\rho_0u_0\approx \frac{2F_{c0}}{v_{esc}^2\cdot A_{gal}},
\end{equation}\label{massloss_final}

\noindent where $F_{c0}$ denotes the spatially averaged energy flux
density of the cosmic rays and $A_{gal}$ is the surface area of the
galactic disc.

This model is ideally suited for high resolution N-body/hydrodynamic
simulations, in which the galaxies and the inter-stellar medium
(ISM) are spatially resolved. Most quantities in Eq.
\ref{bernoulli_3} would be known directly from the simulations and
the initial outflow velocity $u_0$ could be directly measured in
such simulations. Our setup, however, does not allow such a
procedure, as we do not have the internal velocity field of the ISM
near the reference level. Therefore, in the following, we have to
rely on several assumptions about the thermodynamic properties of
the ejected gas, i.e. its temperature and density. These
uncertainties have to be kept in mind when interpreting the
presented results.

\section{Simulations}

We implement the prescription of galactic wind mass loss given by
Eq. \ref{massloss_total} into the setup described in Kapferer et al.
(2007a) which was improved with respect to the setup used in
Schindler et al. (2005), Kapferer et al. (2006) and Domainko et al.
(2006) by the use of comoving coordinates in the hydrodynamic code.
In this setup we use different code modules to calculate the main
components of a galaxy cluster in the framework of a standard
$\Lambda$CDM cosmology, i.e. using $\Omega_{\rm m}=0.3$,
$\Omega_\Lambda=0.7$, $h=0.7$, $\sigma_8=0.93$, and $\Omega_{\rm
b}=0.02 h^{-2}$. In the following subsections we will detail these
code modules and their interplay.

\subsection{Dark Matter and galaxy formation models}

The non-baryonic dark matter (DM) component is calculated using
GADGET2 (Springel 2005) with constrained random field initial
conditions (Hoffman \& Ribak (1991), implemented by van de Weygaert
\& Bertschinger (1996)). From the resulting DM distribution we
calculate at each time step the underlying gravitational potential
for the hydrodynamic code by solving the Poisson equation.
Additionally we can calculate the dynamically fully described orbits
and the merger tree of galaxy haloes from the DM distribution.

The properties of the galaxies are calculated by an improved version
(van Kampen et al. 2005) of the galaxy formation code of van Kampen
et al. (1999) which is a semi-analytic model in the sense that the
merging history of galaxy haloes is taken directly from the
cosmological N-body simulation. Halo mergers are identified in the
N-body simulations as events in which a new halo is formed that is
both virialised and contains at least two progenitors. Gas that
cools within virialised haloes is assumed to settle in a disc with
an exponential profile. To set the disk-scale length we use the
model of Mo, Mao \& White (1998), with the distinction that we
measure the angular momentum of the dark halo from the N-body
simulation data instead of assuming an analytical model for it.
Gas-rich galaxies undergo a burst of star formation during a halo
merger event. Galaxy-galaxy mergers are treated in a different way:
all galaxies within a dark matter halo or subhalo suffer from
dynamical friction, and will merge with the central galaxy
eventually, triggering a starburst as well. This starburst uses up
most of the cold gas left in the satellite, however not all is
turned into stars due to supernovae feedback, which heats the
surrounding gas. Stars which are formed during halo merger events
and galaxy-galaxy mergers contribute to the stellar mass of the
bulge. In between merger events star formation is quiescent in discs
with a threshold according to the Kennicutt criterion (Kennicutt
1998). Stellar evolution is modelled using the stellar population
synthesis models of Bruzual \& Charlot (2003). A fraction of the
metals formed in stars is ejected by supernovae and stellar winds
into the surrounding inter-stellar medium (ISM). The metals that end
up in the hot gas component also affect the metallicity dependent
cooling rates (Sutherland \& Dopita 1993). Yields are taken from the
metallicity evolution model of Matteucci \& Francois (1989), in a
manner as described in van Kampen et al. (1999), which includes SN
types I and II. The evolution of the metallicity is followed for the
stellar populations as well as for the cold and hot gas reservoirs,
and exchanges are tracked as well. Material returned to the cold ISM
by stellar winds and supernovae has been chemically enriched by the
nuclear processes inside the stars.

\subsection{Hydrodynamics}

For the treatment of the ICM we use comoving Eulerian hydrodynamic
parts with shock capturing scheme (PPM, Collela \& Woodward 1984),
with a fixed mesh refinement (Ruffert 1992) on four levels and
radiative cooling (Sutherland \& Dopita 1993). For the initial
conditions of the hydrodynamic simulation we distribute the gas as
the dark matter in a Gaussian random field at $z=40$. The largest of
the four nested grids covers a comoving volume of (20 Mpc)$^3$. Each
finer grid covers $\frac{1}{8}$ of the volume of the next larger
grid. Using a resolution of 128$^3$ grid cells on each grid we
obtain a finest spacing of $\sim$19.5 kpc comoving for each cell on
the innermost grid. The N-body and hydrodynamic code are calculated
starting at $z=40$, while the semi-analytic galaxy formation model
covers the redshift interval from $z=20$ to $z=0$.

\subsection{The model galaxy clusters}

With this numerical setup we model a sample of merging systems, from
which we draw a non-cooling flow cluster which is a lower mass
cluster with several strong sub-cluster mergers. We will name this
model cluster hereafter 'model cluster A'. The second model cluster
we use for this analysis is a more massive merging system with about
twice the mass of model cluster A and several minor merger events.
This model cluster will be referred to as 'model cluster B'. These
model clusters were also used in Kapferer et al. (2007a), with
respect to the initial conditions of the dark matter and the
baryonic matter. Main improvements were done regarding the
semi-analytical galaxy formation model. One improvement is the
change of the IMF, where we now use the IMF proposed by Chabrier
(2003). This produces better fits to the luminosity function without
large changes to the star formation history. Other improvements
include adopting a 40\% shorter galaxy merger rate timescale, and a
different ratio for the distinction between minor and major mergers,
which is now 0.3. In the previous model the metallicity of the cold
gas in the disc galaxies was higher, therefore the stripping leads
to higher metallicities in the ICM. To constrain the model
parameters for the semi-analytical model which is applied, we made
sure to reproduce the Tully-Fisher relation and the global star
formation history. The luminosity functions in the B-band (star
forming galaxies) and the K-band (passively evolving galaxies) are
in good agreement around $L_*$ in both bands, but overpredict
abundances at the bright end. This problem is usually fixed by
implementing an AGN feedback recipe, which the model used for this
work does not include. For completeness we summarise here the main
properties of the two model clusters as described in Kapferer et al.
(2007a).

\begin{itemize}

\item \textbf{Model Cluster A:} The cluster starts to form at z$\sim$1.5
and has two major merger events at z=0.8 and z=0.5. The total mass
at z=0 is $1.5\times10^{14}$ M$_{\odot}$ in a sphere with 1 Mpc
radius.

\item \textbf{Model Cluster B:} The formation redshift for this
cluster is z$\sim$1.7. Four minor merger events appear at z=1.4,
z=1.1, z=0.5 and z=0.3. The cluster has a final mass of
3.4$\times10^{14}$ M$_{\odot}$ within a sphere of 1 Mpc radius.

\end{itemize}

\subsection{Implementation of the analytical wind model into the numerical setup}

The semi-analytic galaxy formation model limits the applicability of
Eq. \ref{bernoulli_3}, because it does not give spatially resolved
information about the star formation. Instead it gives a global
value for the disc. Density and temperature distributions are also
unknown. Therefore, to make use of Eq. \ref{bernoulli_3} in our
numerical setup, we have to rely on several assumptions. To
calculate the gravitational potential and its escape velocity, we
assume for the dark matter halo a Hernquist (1990) potential:

\begin{equation}\label{hernqu_pot}
\Phi_{DM}(r)=-\frac{GM_{DM}}{r+a},
\end{equation}

\noindent where $a$ is a scale length and M$_{DM}$ the total mass of
the dark matter halo. The density profile associated with this
potential via the Poisson equation reads

\begin{equation}\label{hernqu_rho}
\rho_{DM}(r)=\frac{M_{DM}}{2\pi}\frac{a}{r(r+a)^3}.
\end{equation}

For the potential of the disc we use a Toomre (1963) model (first
derived by Kuzmin (1956)), which is given by

\begin{equation}\label{toomre_pot}
\Phi_{disc}(r,z)=-\frac{GM_{disc}}{[r^2+(b+|z|)^2]^{1/2}},
\end{equation}

\noindent where $r$ and $z$ denote the cylindrical coordinates and
$b$ is a non-zero scale length. The potential at a certain radius
can then be directly calculated from Eqs. \ref{hernqu_pot} and
\ref{toomre_pot}, using the total mass of the disc and of the dark
matter halo. The scale length $a$ in Eq. \ref{hernqu_pot} follows
from the disc scale length of an exponential disc $r_d$ as
$a=r_d/(1+\sqrt{2})$ (Hernquist 1990).

The terminal velocity of the outflow $u_{\infty}$ is also unknown,
so we assume $u_{\infty}=v_{esc}$. Breitschwerdt et al. (1991) have
shown that the terminal velocity of the wind $u_{\infty}$ and the
escape velocity of the system $v_{esc}$ differ typically by not more
than a factor of two. We have tested the consequences of this
uncertainty by using $u_{\infty}=v_{esc}/2$ and
$u_{\infty}=2v_{esc}$, respectively. We found that quantitatively
the resulting absolute metallicity values change up to a factor of
three. The resulting metal distribution (radial profiles and 2D
metal maps) and the redshift evolution of the metallicity, however,
are only marginally affected. Given all the uncertainties in the
simulation setup, e.g. within the semi-analytical galaxy formation
model, this uncertainty appears acceptable.

To derive the sound speed we would require a pressure and density
distribution through the gaseous disc, which is not available from
the semi-analytic model. Therefore, we follow the argumentation line
of Breitschwerdt et al. (1991), who estimated that for a Milky-Way
type galaxy with a supernova rate of once every 30 years the time
averaged halo filling factor is of order unity. Therefore the
pressure support for a flux tube should be, on average, maintained.
We couple the supernova rate to the star formation rate (SFR) via
the assumption of a Salpeter initial mass function (IMF) with a
slope of -1.35 in the range of 0.1 M$_{\odot}$ and 40 M$_{\odot}$. A
model galaxy is hence having a galactic wind driven purely by
thermal pressure if the SFR exceeds a certain threshold, for which
we adopt here a value of 4 M$_{\odot}$/yr. Below this SFR the galaxy
can still have a cosmic ray driven wind, for which we calculate the
mass loss using Eq. 10. As in Breitschwerdt et al. (1991) we then
assume that the flux tubes are homogeneously distributed over the
galactic disc. This obvious oversimplification is necessary, as we
have no information on the location of the supernovae from the
semi-analytical model. We set the temperature and the density at the
reference level to 1$\times 10^{7}$ K and 1$\times 10^{-26}$
g/cm$^3$, respectively. These are typical values for gas fed into a
galactic wind (e.g. Breitschwerdt et al. 1991). For an outflow to
occur we furthermore demand that the pressure of the outflow exceeds
the pressure of the surrounding ICM.

\section{Results}

In the following we investigate the efficiency and the time
dependency of galactic winds as enrichment process. In order to
compare our results to observed quantities we will present mean
metallicities, metallicity profiles, and 2D metal maps of our model
clusters. In addition we compare the efficiency of the proposed wind
model with ram-pressure stripping as source of metals in the ICM.
Metals originating from galaxies, either from galactic winds or
ram-pressure stripping, are advected and mixed with the hydrodynamic
solution of the ICM for each timestep in the simulation.

\subsection{Metallicity profiles}\label{section_profile}

\begin{figure}
\begin{center}
{\includegraphics[width=10cm]{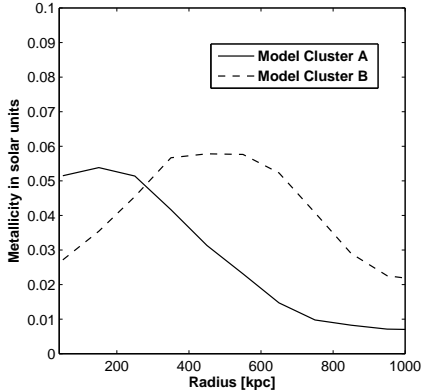}} \caption{Radial
metallicity profile of model cluster A and B at redshift $z=0$. For
the shown metallicity profiles only metals originating from galactic
winds are taken into account.} \label{profile}
\end{center}
\end{figure}

The most basic measure to quantify the metal distribution in the ICM
is its mean metallicity, obtained from emission weighted metallicity
maps. We assume that the X-ray measurements extend to a radius of 1
Mpc and that only galactic winds are acting. Within this volume, the
mean metallicity in solar units (Grevesse \& Sauval 1998) is 0.02
and 0.04 at $z=0$ for model cluster A and B, respectively. For both
model clusters this value is significantly lower than observed
metallicities in real clusters. More information on the metal
distribution is given by radial metallicity profiles, especially by
their slope. The normalisation of the profile depends strongly on
the uncertain ISM metallicity, present in semi-analytic simulations.
In Fig. \ref{profile} we present the metallicity profile of model
cluster A and B at redshift z=0. Note that for these profiles only
metals originating from galactic winds are taken into account. The
metal profiles are obtained from X-ray weighted metal maps (see
Kapferer et al. 2007b). For model cluster A the metallicity is
almost constant up to a radius of $\sim$300 kpc. Beyond $\sim$500
kpc the profile decreases constantly. Model cluster B is more
massive than model cluster A and, as a result of a prominent infall
region, the metallicity at a radius of 1 Mpc is with $\sim$0.03
solar units higher than in cluster A. A striking feature of the
profile is the decreasing metallicity towards the centre of the
cluster. This behaviour is different from model cluster A. Galactic
winds are suppressed by the high central pressure of the ambient
medium, a phenomenon which was already discussed by Schindler et al.
(2005) for quiet (i.e. mainly cosmic-ray driven) galactic winds.

\begin{figure}
\begin{center}
{\includegraphics[width=\columnwidth]{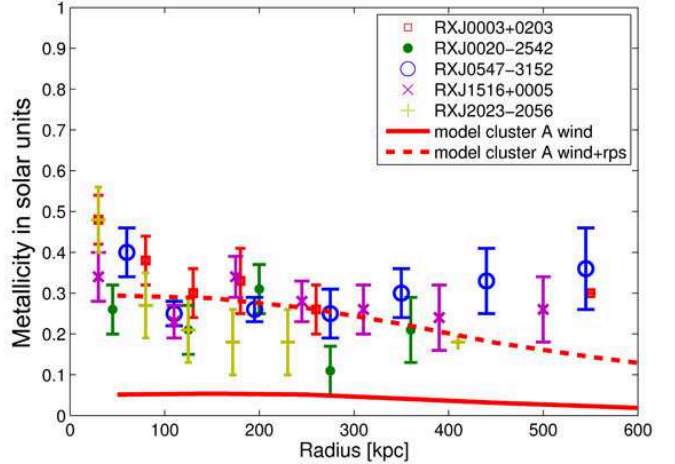}}
\caption{Metallicity profiles for observed nearby ($z<0.2$) clusters
(symbols) from Pratt et al. (2007) compared to emission weighted
simulated profiles from model cluster A at redshift $z=0$ with
galactic winds (red, solid line) and with combined wind and
ram-pressure stripping (red, dashed line) as enrichment processes.}
\label{comp_obs_602}
\end{center}
\end{figure}

Finally, we compare the simulated metallicity profiles to observed
profiles from Pratt et al. (2007) in Fig. \ref{comp_obs_602}. Using
only galactic winds as enrichment process (red, solid line), the
radial metallicity profile of model cluster A lies at all radii
significantly below the observed profiles. The slope agrees
reasonably well with the observed values. This result indicates,
that additional processes are acting in clusters, presumably
ram-pressure stripping and/or AGNs. We calculate the combined effect
of galactic winds and ram-pressure stripping and present the result
in Fig. \ref{comp_obs_602} (red, dashed line). Ram-pressure
stripping was modelled using a Gunn \& Gott (1972) criterion, as
described in Kapferer et al. (2007a). The profile agrees then
reasonably well at most radii. In the very centre (the inner 50
kpc), where our resolution is not sufficient, a central AGN might
still be responsible for the observed steepening there. In the very
outer parts of the cluster the simulated metallicity profile lies
below the observed values.

\begin{figure}
\begin{center}
{\includegraphics[width=\columnwidth]{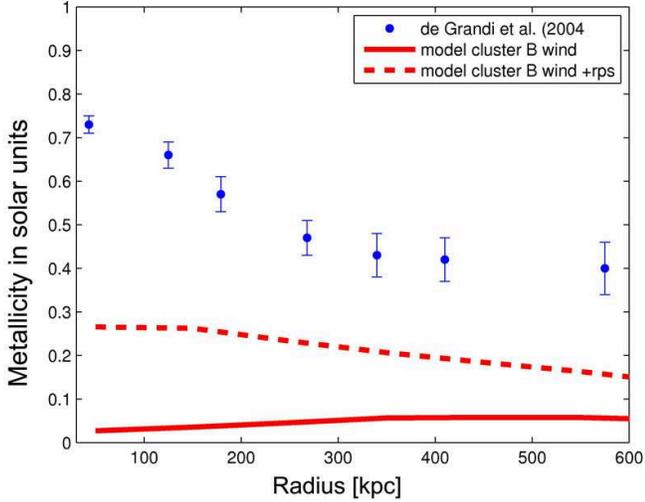}} \caption{Mean
metallicity profile from a sample of observed clusters (symbols)
from de Grandi et al. (2004) compared to emission weighted simulated
profiles from model cluster B with galactic winds (red, solid line)
and with combined wind and ram-pressure stripping (red, dashed line)
as enrichment processes.} \label{comp_obs_604}
\end{center}
\end{figure}

We carried out the same analysis for model cluster B. However, as
this model cluster is mostly relaxed at redshift $z=0$, it is not
reasonable to compare it to the set of merging clusters from Pratt
et al. (2007). Therefore we use the mean of cool-core clusters
presented in De Grandi et al. (2004) for comparison. The result is
shown in Fig. \ref{comp_obs_604}. Again, with wind as the only
enrichment process, neither the slope nor the absolute values agree
with the observations, especially in the inner parts of the cluster.
Combining winds and ram-pressure stripping leads to reasonable
agreement in the slope of the profiles for radii $r\geq120$kpc. At
smaller radii, however, the mean observed profile is steeper than
our simulated cluster profile. This might be a selection effect, as
we show only a snapshot of one specific model cluster. It may also
indicate, however, that we miss an enrichment process acting in the
centre of massive, relaxed clusters (e.g. enrichment by a central
AGN).\\
In the case of the massive galaxy cluster the difference between the
phenomenological model for galactic outflow in Kapferer et al.
(2007a) and the present work is more striking. In the
phenomenological wind model the contribution to the enrichment of
the ICM is up to a factor of ten more efficient compared to the wind
model presented in this work. The analytical wind model introduced
in this work gives much less metallicities compared to the outflow
model which scales the outflow linearly with the galaxies' star
formation rate. The strong mismatch between the reproduction of the
observed metallicity profiles in the case of the massive system B
leads to the conclusion that additional sources for metal enrichment
in massive galaxy clusters play an important role. In addition the
amount of metals originating from ram-pressure stripping is much
lower compared to Kapferer et al. (2007a). The explanation lies in
the improved semi-analytical model, see section 3.3. The
investigation of the metal enrichment through AGNs will be part of
an upcoming work.

We note once more, that the absolute values of the profile depend
strongly on the adopted parameters of the semi-analytic galaxy
formation model and the parameters of the wind model. The slope of
the simulated profile, on the contrary, is robust against reasonable
changes of the parameters.

Similar studies have been performed by other groups. Tornatore et
al. (2007), for example, used an N-body/SPH code. They found
generally good agreement of the metallicity profiles with
observations. The same is true for Cora (2006), whereas they apply a
semianalytical model in their work. Their adopted wind model might,
however, overestimate the mass loss caused by galactic winds (see
Sect. 4.3.).

\begin{figure}
\begin{center}
{\includegraphics[width=\columnwidth]{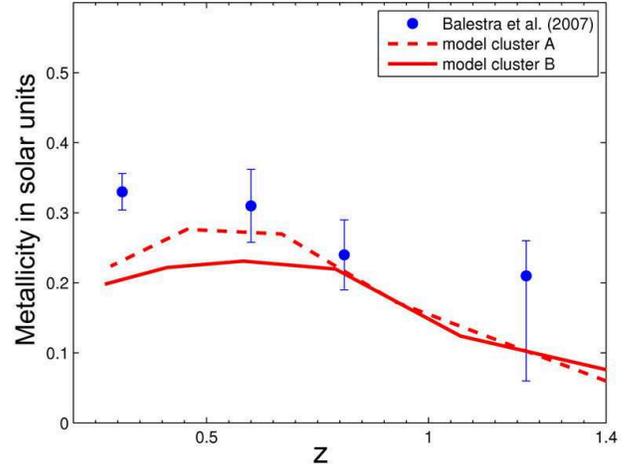}}
\caption{Redshift evolution of the mean metallicity of model cluster
A (red, dashed line) and model cluster B (red, solid line) compared
to the observed evolution of a sample of clusters from Balestra et
al. (2007). Both enrichment processes (ram-pressure stripping and
galactic winds) are taken into account.)} \label{m_z}
\end{center}
\end{figure}

In Fig. \ref{m_z} we present the redshift evolution of the mean
metallicity of the two model clusters and compare them with the
observed evolution of a sample of clusters from Balestra et al.
(2007). Both enrichment processes (ram-pressure stripping and
galactic winds) are taken into account. For both clusters the
evolution agrees with the observational trend but the absolute
values are lower than expected from observations. This is especially
true at lower redshift (z$<$0.5), where the enrichment processes do
not seem to be efficient enough. In the observational sample of
Balestra et al. (2007) the extraction radius varied roughly from 200
- 1000 kpc, depending on the signal-to-noise ratio. We assumed a
constant extraction radius of 600 kpc, which we also use for
comparison of the metallicity profiles with observations. A
variation of this radius does not change the general trend
significantly. Note that 1 Mpc corresponds roughly to the virial
radius of the two galaxy clusters at $z=0$.

\subsection{X-ray weighted maps}

\begin{figure}
\begin{center}
{\includegraphics[width=\columnwidth]{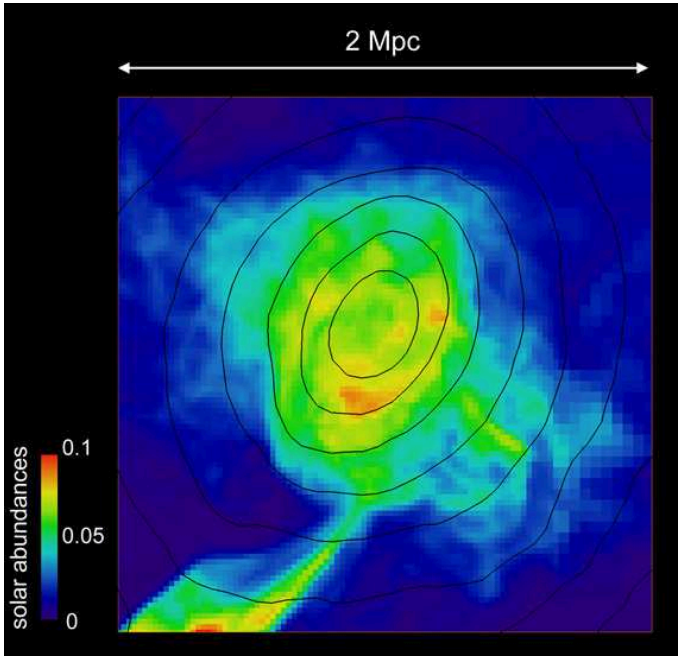}}
\caption{X-ray weighted metal map for model cluster A at redshift
z=0. Overlayed contours indicate the X-ray surface brightness.}
\label{map_602}
\end{center}
\end{figure}

With modern X-ray space telescopes it is feasible to study also the
2D metal distribution. It was found that the metals are not
homogeneously distributed within the ICM (e.g. Sanders et al. 2004;
Durret et al. 2005; O'Sullivan et al. 2005; Sauvageot et al. 2005;
Werner et al. 2006; Sanders \& Fabian 2006; Hayakawa et al. 2006;
Finoguenov et al. 2006; Bagchi et al. 2006). In Fig. \ref{map_602}
we present such an X-ray weighted metal map for model cluster A at
redshift z=0. The X-ray surface brightness map is overlayed as
contours and shows the overall ICM distribution. A filament along
which galaxies fall into the cluster is visible at the bottom left.
As the ICM density in the filament is relatively low
($\sim$7$\times$10$^{-29}$g/cm$^3$ compared to
$\sim$7$\times$10$^{-27}$g/cm$^3$ in the centre), it can be more
easily enriched there (i.e. a higher metal mass fraction is
achieved). In the centre of the cluster the metallicity is slightly
lower.

\begin{figure}
\begin{center}
{\includegraphics[width=\columnwidth]{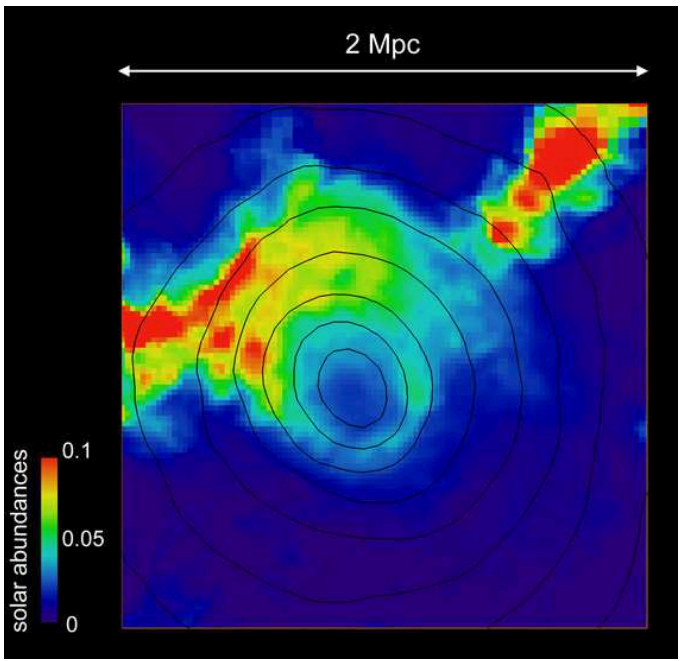}}
\caption{X-ray weighted metal map for model cluster B at redshift
z=0. Overlayed contours indicate the X-ray surface brightness.}
\label{map_604}
\end{center}
\end{figure}

In Fig. \ref{map_604} we present the X-ray weighted metal map for
model cluster B at redshift z=0. The X-ray surface brightness map is
again overlayed as contours and shows the overall ICM distribution.
The suppression of the galactic winds in the centre of the cluster,
as discussed in Sect. \ref{section_profile}, is clearly visible. In
the left and upper-right parts of the cluster the outflows of
galaxies, falling into the cluster along a filament, cause a complex
metallicity pattern with several maxima.

\section{Summary and conclusions}

We have investigated the efficiency and time-dependence of thermally
and cosmic ray driven galactic winds for the metal enrichment of the
intra-cluster medium (ICM) by a combination of a new analytical
model and cosmological hydrodynamic simulations. The analytical
approximations for the mass loss by galactic winds driven by thermal
and cosmic ray pressure were derived from the Bernoulli equation. We
extend in this work the previous wind model presented in Kapferer et
al. (2007a) by introducing a new analytical model for the mass loss
due to galactic winds. This new model does allow a deeper insight
into the physical processes involved in the outflow and addresses
environmental effects such as the possible suppression of galactic
winds by high ICM pressure (Schindler et al. 2005). It is still
simple enough to be readily implemented in metal enrichment and
semi-analytic galaxy formation models. We simulated the enrichment
processes for a galaxy cluster with major merger events (cluster A)
and a less merging system (cluster B). Investigating observable
quantities, we found that:

\begin{itemize}
\item Galactic winds alone cannot account for the observed mean metallicity of the ICM.
At redshift $z=0$ the model clusters have mean metallicities, which
are almost a factor of 10 lower than observed values. Within a
radius of 1 Mpc model cluster A has a mean metallicity of 0.02 and
model cluster B a metallicity of 0.04 in solar units.

\item A central drop in the metallicity of the relatively relaxed model
cluster B is due to a suppression of steady state galactic winds by
the pressure of the ambient ICM. In the merging model cluster A, no
decrease of the metallicity towards the centre is observable.

\item 2D metallicity maps reveal important additional information on the metal
distribution, e.g. infalling groups from filaments and
inhomogeneities.

\item Combining ram-pressure stripping and galactic winds we
find radial metallicity profiles of the model clusters which agree
in their slopes with observations. In the inner parts of massive
clusters the observed profile is steeper than in the simulations.

\item The slope of the redshift evolution of the mean metallicity in the
simulations agrees reasonably well with recent observations of
Balestra et al. (2007).

\end{itemize}

\subsection{What additional processes do enrich the ICM}

Ram-pressure stripping and galactic winds, as modelled in this work,
are not able to account for the observed metallicities in the ICM.
Therefore additional processes need to be taken into account. An
important source of enriched matter in the central part of galaxy
clusters would be an AGN in a massive central cluster galaxy. The
injection of buoyant bubbles in the ICM transports enriched matter
from the central cluster galaxy to the surrounding ICM (Roediger et
al. 2007). Simionescu et al. (2009) find that 15\% of the total
amount of Fe in the inner region of the galaxy cluster Hydra A
belongs to gas, which is uplifted by buoyant bubbles from the
central AGN. This effect would be stronger in relaxed, more massive
systems, as the stratification of the ICM is not strongly disturbed
by mergers and therefore the bubbles can survive longer.\\
Ram-pressure stripping, as it is able to trigger star formation in
the gaseous wake of stripped galaxies, can enrich the ICM in situ
(Kronberger et al. 2008, Kapferer et al. 2008, Kapferer et al.
2009). In the case of strong ram pressures stripping nearly all new
stars are formed in the wake of the stripped galaxies. Their
influence on the overall metal enrichment is yet unknown, but first
simulations of ram-pressure stripping including recipes for star
formation point towards the picture that off disc star formation due
to ram-pressure stripping is very efficient as an enrichment process
of the ICM. Also Intra-cluster stellar populations
can enrich the ICM very efficiently as shown in Sivanandam et al.(2009). \\
Galaxy-galaxy interaction is able to enrich the surroundings by
redistributing the ISM kinematically in large volumes around the
interacting galaxies (Kapferer et al. 2005). In the model of
galactic winds, described in this work, the effect of mergers on the
star formation are taken into account, which lead to stronger
outflows in the galactic winds. Nevertheless the redistribution of
the ISM is not modelled within the presented simulations. The
heating of the kinematically redistributed ISM, so that it can
contribute to the observed X-ray flux, is not well understood. Hence
conclusions on the contribution to the overall
enrichment of the ICM due to galaxy interactions have to be investigated in detail.\\
In a recent paper Fabjan et al. (2008) studied the redshift
evolution of the metal content of the ICM with N-body/SPH
simulations combined with a detailed model of chemical evolution.
They found good agreement with observations for a Salpeter initial
mass function. They have shown that gasdynamical processes which
redistribute metals, like the sinking of low entropy enriched gas to
the central cluster regions, is an additional process that helps
explain the increase of the observed emission-weighted metallicity
of the ICM at low z. These results also confirm those found by Cora
et. al (2008) which were obtained by using a combination of
N-Body/SPH simulations with an analytic model of galaxy formation.\\
Concluding, the effects of AGN, ram-pressure stripping induced
off-disc star formation and direct enrichment by galaxy-galaxy
interactions will be investigated in upcoming works, with the aim to
obtain a more complete picture of the relative strength of the
enrichment processes in galaxy clusters and the influence on the
evolution of the galaxies.

\begin{acknowledgements}
We thank the anonymous referee you helped to improve the quality of
the paper significantly. The authors further acknowledge the
UniInfrastrukturprogramm des BMWF Forschungsprojekt Konsortium
Hochleistungsrechnen, the ESO Mobilit\"atsstipendien des BMWF
(Austria), the Austrian Science Foundation (FWF) through grants
P18523-N16, P18416-N16, P18493-N08, and P19300-N16, and the Tiroler
Wissenschaftsfonds (Gef\"ordert aus Mitteln des vom Land Tirol
eingerichteten Wissenschaftsfonds). We also thank Rien van de
Weygaert and Ed Bertschinger for allowing us the use of their
constrained random field code and Volker Springel for GADGET-2.
\end{acknowledgements}

\end{document}